\documentclass[pdflatex,sn-nature]{sn-jnl}

\usepackage{amsmath,amssymb,amsfonts,mathtools}
\usepackage{siunitx}
\usepackage{bm}
\usepackage{booktabs, multirow, adjustbox}
\usepackage{graphicx}
\usepackage{float}
\usepackage{algorithm}
\usepackage{algpseudocode}
\usepackage{enumitem}
\usepackage{amsthm}
\usepackage{newunicodechar}
\newunicodechar{∼}{\textasciitilde}
\usepackage[title]{appendix}
\usepackage{textcomp}
\usepackage{multirow}
\usepackage{subcaption}
\usepackage{hyperref}
\usepackage{xcolor}
\usepackage{xurl}

\usepackage[dvipsnames,svgnames,x11names]{xcolor}
\definecolor{DCTop20}{HTML}{E69F00}
\definecolor{DCFeasible}{HTML}{0072B2}
\definecolor{DCTransmission}{HTML}{B9BFC6}

\usepackage[utf8]{inputenc}

\theoremstyle{thmstyleone}

\theoremstyle{thmstyletwo}

\theoremstyle{thmstylethree}

\newcounter{boxcounter}
\renewcommand{\theboxcounter}{\arabic{boxcounter}}

\unnumbered

\begin{document}

\title{Flexibility-Aware Framework for Efficient Planner-Initiated Siting of Data Center}

\author[1]{\fnm{Dongjoo} \sur{Kim}}
\author[1]{\fnm{Lin} \sur{Dong}}
\author*[2]{\fnm{Le} \sur{Xie}}\email{xie@seas.harvard.edu}

\affil[1]{\orgdiv{Department of Electrical and Computer Engineering}, \orgname{Texas A\&M University}, \orgaddress{\city{College Station}, \postcode{77843}, \state{TX}, \country{USA}}}
\affil[2]{\orgdiv{Harvard John A. Paulson School of Engineering and Applied Sciences}, \orgname{Harvard University}, \orgaddress{\city{Allston}, \postcode{02134}, \state{MA}, \country{USA}}}

\abstract{
Explosive growth in energy-intensive AI data centers is outstripping the pace of power grid interconnection and transmission expansion. While operational flexibility has been proposed to mitigate this stress, existing processes are often reactive and evaluate projects only after they enter a multi-year interconnection queue. To address this, we introduce a planner-initiated siting framework that integrates (i) reliability-gated screening, (ii) system-wide market-impact assessment under standardized flexibility envelopes (firm, pause, and shift), and (iii) entropy-weighted multi-criteria scoring to produce ranked, pre-certified catalogues of interconnection-ready locations. Applied to a synthetic 2,000-bus Texas power system, the framework demonstrates that operational flexibility expands the siting frontier by 9–17\% at 1 GW and 19–21\% at 2 GW compared to firm operation. Median all-hour average prices remain essentially unchanged (\$24.32/MWh for the 2 GW cases), and the shift envelope attenuates peak-hour price dispersion by approximately 3.4\% with minimal side effects during off-peak hours. Utilizing pre-certified envelopes to bypass major transmission reinforcements, this workflow enables first energization in 12–18 months—a conservative reduction of 3.5–4 years versus the conventional 5–8 year project-led process. This technology-agnostic framework provides a proactive decision-making tool for system operators and regulators to fast-track large flexible loads while preserving grid reliability and market stability.}


\maketitle

\section{Introduction}\label{sec:intro}
Within one decade, the world’s digital infrastructure  could require the electricity to the equal of major industrial nations. Data centers used approximately \SI{460}{\tera\watt\hour} in 2022 and are projected to approach \SI{0.95}{\peta\watt\hour} by 2030—about \SI{3}{\percent} of global electricity—with AI-optimized workloads the main driver \cite{IEA2025}. The speed and geographic concentration of this growth, rather than growth alone, is now challenging power systems.

The efficiency-kept equilibrium of the 2010s has ended; AI-scale compute is driving sustained load growth that outpaces interconnection and grid build-out, creating a rapidly widening gap between digital infrastructure deployment and the observed pace of power-system expansion \cite{Masanet2020,BCG2025,JLL2025}. This growing chasm motivates the central premise of this research. In North America, regional transmission organization (RTOs) explicitly revise load trajectories to account for unprecedented data-center–driven growth: the Pennsylvania-New Jersey-Maryland Interconnection (PJM) adds zone-level adjustments (including Dominion, VA), lifting the ten-year RTO peak by tens of gigawatts \cite{PJM_2024_LoadForecast}, while Virginia’s legislative audit finds that unconstrained demand could double within a decade, requiring substantial new in-state generation and a \(\sim\)40\% increase in interzonal transfer capability into Dominion \cite{JLARC_2024_DataCentersVA}. Meanwhile, global grid stocktakes and federal U.S. studies conclude that transmission development times and connection queues have become binding constraints, with pervasive regional and interregional needs emerging by 2030 and beyond \cite{IEA_Grids_2023,DOE_Transmission_Needs_2023}. In practice, data center campuses can be site-ready in \(\sim\)2 years, whereas major substation or transmission augmentations often require \(\sim\)5–10 years—an asymmetry that fuels interconnection queues, late reinforcements, and localized price volatility \cite{NERC2023,NERC2024,NERC2025,ey2025datacenter,Saul2024BloombergLawDominion}.

Although data-center loads often exhibit relatively simple and stable demand profiles—characterized by flat, highly predictable electricity consumption over short time horizons—they are structurally distinct from most other forms of load growth from a planning perspective. Hyperscale data centers introduce large, persistent, and spatially concentrated demand at a limited number of candidate interconnection points, placing sustained stress on generation adequacy and local transmission capacity and tightening reliability constraints due to their low tolerance for supply interruptions. These challenges are amplified by pronounced time-scale mismatches and demand uncertainty: data-center projects can become site-ready within one to two years, whereas major transmission or substation reinforcements typically require much longer planning and construction horizons, and future demand may scale rapidly, relocate, or be withdrawn. At the same time, data-center electricity demand increasingly exhibits structured and contractible forms of flexibility enabled by computational workload scheduling, such as temporal load shifting or managed curtailment.

Recent studies and regulatory initiatives have explored alternatives to traditional project-led interconnection, including hosting-capacity analyses, expedited interconnection pathways, non-firm or flexible connections \cite{ERCOT2022InterimNotice, cicilio2020transmissionhotsing, bjarghov2024enhancingnonfirm}. While these approaches represent important progress, they are often limited to specific regions, single planning stages, or isolated technical criteria. In practice, the absence of a system-level, planner-initiated workflow that jointly considers reliability, operational flexibility, and market impacts has led to fragmented decision-making and continued reliance on reactive queue-based processes.

Despite emerging utility- and regulator-led initiatives, practice remains piecemeal and lacks an integrated, auditable workflow that links reliability-gated screening, standardized flexibility envelopes, and market-impact quantification—gaps this study addresses. Signals of both urgency and fragmentation are evident across jurisdictions. Northern Virginia’s Data Center Alley (\(\sim\)\SI{3.6}{\giga\watt} coincident demand in 2024) has triggered transmission reinforcements and contentious local siting debates \cite{NESCOE2024}. Internationally, Ireland has proposed a dedicated connection policy for large energy users; Singapore paused then relaunched capacity via a Green Data Centre Roadmap that unlocks at least \SI{300}{\mega\watt} subject to efficiency and clean-power criteria.; and the Netherlands temporarily halted hyperscale projects pending national rules \cite{CRU2025,IMDA2024,NL2022}. Great Britain’s Connections Action Plan documents customers receiving dates in the late 2030s under status quo processes and sets out queue management and non-firm/flexible connection reforms \cite{DESNZ_Ofgem_Connections_2023}. In the U.S., Electric Reliability Council of Texas (ERCOT) has formalized a Large Load Interconnection Study process, and the Southwest Power Pool (SPP) has floated an expedited \(\sim\)90-day fast-track concept for qualifying data centers \cite{ERCOT2025LLIS,Plautz2025EENewsSPP}.

These developments suggest a different organizing principle for large-load interconnection: a shift from project-led queueing toward a planner-initiated siting workflow in which interconnection feasibility is treated as the primary design constraint and operational flexibility is evaluated as a standardized planning instrument. In this emerging workflow, planners pre-define operating envelopes at candidate nodes, screen them for reliability compatibility, and assess system-level effects before queue entry, thereby producing auditable shortlists that are consistent with ongoing reforms around flexible and fast-track interconnection \cite{Ghatikar2010, AlKez2021,Takci2025,CamusEnergy2025}.

In this work, we introduce a planner-oriented computational framework for the proactive, reliability-gated, and market-driven siting of large, potentially flexible loads. The framework represents flexibility through standardized operating envelopes (firm, pause, and shift), screens candidate locations for compatibility with N--1 security, quantifies system-wide market effects using transparent price and congestion metrics, and ranks feasible sites using a reproducible multi-criteria procedure. The contribution is architectural in primacy but not exclusively so: the workflow is underpinned by (i) formally parameterized flexibility envelopes as contractible planning instruments enabling pre-certification and ex post audit; (ii) simultaneous, system-scale coupling of N--1 reliability gating with SCUC/SCED-based market simulation across multiple operating envelopes, distinct from existing project-by-project hosting-capacity and interconnection studies; and (iii) entropy-weighted TOPSIS applied to grouped, direction-consistent absolute market metrics, a combination not previously applied in the large-load interconnection siting context. Using a synthetic Texas-scale power grid, we show that operational flexibility expands the feasible siting frontier, supports near-term energization pathways, and generally preserves all-hour price levels while reducing peak-time volatility in flexibility-enabled cases. We also identify policy integration points that align the framework with emerging fast-track and flexible-connection processes~\cite{NERC2023,NERC2024,NERC2025,ERCOT2025LLIS}.

\section{Results}\label{sec:results}

\subsection{A planner-initiated, reliability-gated and market-driven framework for fast-track interconnection}\label{sec:framework}

\begin{figure}[hbt!]
\centering
\includegraphics[width=\textwidth]{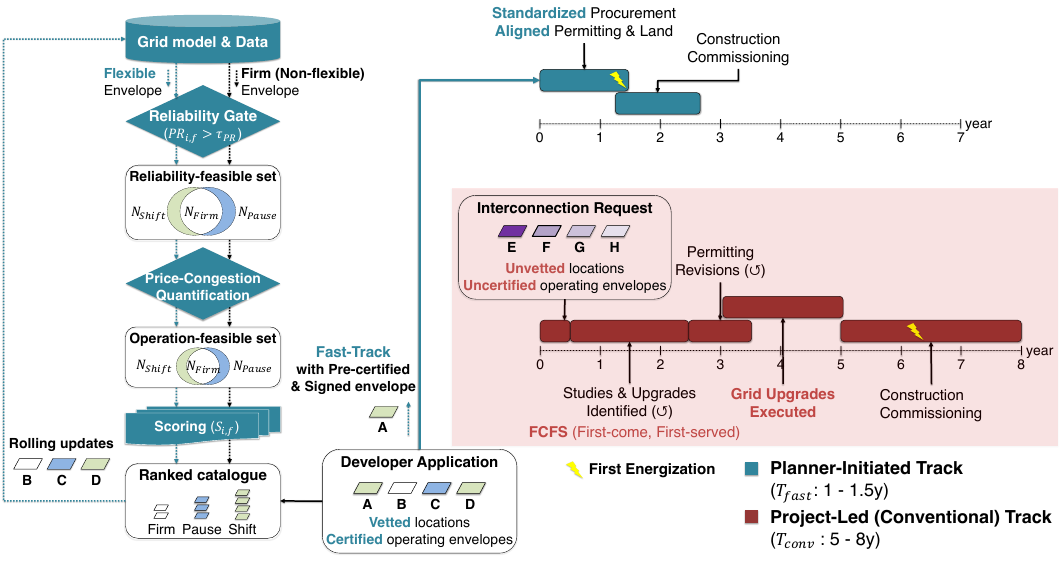}
\caption{Planner-initiated screening-to-certification workflow with indicative timelines.
Starting from a common grid model and data, planners apply a three-stage pipeline comprising Stage~1 reliability gating, Stage~2 price--congestion and operational-feasibility quantification, and Stage~3 scoring and ranking. This process produces (i) a pre-certified ranked catalogue of feasible locations and (ii) site-specific pre-agreed operating envelopes specifying maximum demand, ramp bounds, and optional pause or shift commitments.
A--D denote example developer applications selected from the planner-provided ranked catalogue and submitted under certified operating envelopes. Once the selected envelope is accepted and contractually signed, projects may proceed through a fast-track pathway, supporting earlier first energization, indicated by the lightning icon, on the order of 12--18~months.
For comparison, the conventional project-led pathway begins with unvetted requests under a first-come, first-served process. E--H denote example interconnection requests submitted without prior certification, in which developers independently choose locations and operating modes outside the ranked catalogue and certified envelopes. This pathway typically leads to energization on the order of 5--8~years when substantial new supply and transmission are required \cite{CamusEnergy2025,ey2025datacenter}.
Here, $PR_{i,f}$, $\tau_{PR}$ and $S_{i,f}$ denote the reliability-screening metric, screening threshold and ranking score, respectively; $N_{Shift}$, $N_{Firm}$ and $N_{Pause}$ denote candidate sets under shift, firm and pause operating envelopes; and $T_{fast}$ and $T_{conv}$ denote indicative times to energization for the planner-initiated and conventional pathways, respectively. Durations are illustrative and jurisdiction-dependent.}
\label{fig:framework}
\end{figure}

Building directly on the interconnection timing asymmetry, we present a proactive, planner-initiated siting framework that integrates reliability and market effects from the outset and replaces case-by-case negotiation with a pre-certified catalogue of locations and pre-agreed operating envelopes. Figure~\ref{fig:framework} provides a schematic overview of the framework, while Box~\ref{box:workflow} summarizes its step-by-step algorithmic workflow.

This framework has four elements, including a foundational data layer and a three-stage screening-to-certification pipeline.

\begin{enumerate}[label=(\roman*)]
\item Data preparation and management. This foundational layer aggregates network topology, generator parameters, and time-series profiles for load and renewables. Crucially, we preserve the temporal correlations inherent in power system data, acknowledging its non-independent and identically distributed nature. By maintaining chronological alignment across datasets, we ensure that subsequent optimization stages, such as SCUC, can correctly enforce inter-temporal constraints including generator ramping and minimum up and down times, and can accurately evaluate the economic effects of temporal flexibility mechanisms such as load shifting.

\item Reliability-gated screening. A system-wide feasibility screen based on N--1 security and pass-rate thresholds filters candidate buses before detailed study, focusing scarce planning bandwidth on low-risk locations. Reliability serves as a gate rather than a scoring criterion.

\item Market-impact assessment under standardized flexibility profiles. Each candidate is evaluated for both operational feasibility and market impacts under a transparent, planner-defined profile that specifies maximum demand, ramp bounds, and optional flexibility commitments, such as peak shaving or time shifting. Projects that certify to a given profile become eligible for a fast-track path consistent with emerging non-firm and flexible-connection practices \cite{DESNZ_Ofgem_Connections_2023,CamusEnergy2025,ERCOT2025LLIS}.

\item Transparent multi-criteria ranking. Qualified locations are ranked for each profile using five absolute, system-wide metrics: cross-node mean LMP, cross-node LMP standard deviation, P95--P5 range, binding hours, and congestion rent. We adopt entropy-weighted Technique for Order Preference by Similarity to Ideal Solution (TOPSIS) to obtain data-driven, direction-consistent weights while retaining interpretability~\cite{HwangYoon1981,Entropy2020}. In addition, robust aggregates, such as minimum and expected score across profiles, identify sites that remain high-scoring under multiple operating modes (Methods, Stage~3).
\end{enumerate}

\paragraph{Generalizability and extensibility}
While our demonstration uses only reliability as a gate and market impacts as ranked metrics, the framework is designed to be generalizable and extensible. Additional electrical factors and non-electrical siting factors that materially influence data-center development, such as water consumption and local water availability, acoustic impacts including noise contours and sensitive-receptor buffers, air permitting and emissions, land-use and footprint constraints, construction logistics, and environmental-justice screening, can be incorporated in two modular ways:
\begin{itemize}
    \item As additional gates when regulatory thresholds or community standards must be satisfied, such as maximum allowable decibel levels at parcel boundaries, water-rights limits, or exclusion buffers.
    \item As additional scored criteria when trade-offs are appropriate, such as normalized water intensity (m$^3$/MWh), expected noise-exceedance hours, or construction impact indices, combined with the existing metrics through entropy-weighted TOPSIS or alternative multi-criteria methods (Methods, Stage~3).
\end{itemize}
Because operating envelopes and metrics are defined through open templates, planners can extend the library of profiles, including firm, pause, shift, and ramp-bounded variants, as well as the scoring schema without retooling the pipeline, thereby preserving reproducibility while accommodating jurisdiction-specific priorities.

\paragraph{Output of the planning framework}
The framework outputs (a) a pre-certified catalogue of locations that satisfy reliability gates and (b) a site-specific pre-agreed operating envelope, certified upon signature, that can shorten time to energization by enabling fast-track, non-firm interconnection while long-lead reinforcements proceed \cite{CamusEnergy2025,ey2025datacenter}. Together, these outputs convert the planning process from reactive, case-by-case screening into a proactive mechanism that identifies feasible locations in advance and links them to transparent, pre-defined operating conditions.

\subsection{Study system and flexibility setup}\label{sec:study-setup}
We evaluate the framework on a 2{,}000-node synthetic Texas grid \cite{Birchfield2016} with hourly 2020 demand and renewable profiles and realistic generator cost curves. We adopt the ACTIVSg2000 synthetic Texas system to provide a fully reproducible, ISO-scale demonstration. Detailed operational-grade transmission network models and certain operational datasets are commonly subject to confidentiality restrictions, for example CEII-related limitations, which prevents open release of the underlying cases. Accordingly, quantitative results in this paper should be interpreted as proof-of-concept findings on a realistic synthetic footprint rather than universally calibrated estimates. Security and market assessments are performed under an $N$--1 contingency scope with practicability filters on interconnection voltage level (24--500~kV). Only buses that pass Stage~1 advance to Stage~2 simulations, where infeasible or non-convergent cases are excluded before scoring and ranking.

We apply three standardized operating envelopes: firm (flat load at utilization $\rho$), pause (on-peak curtailment with no make-up), and shift (on-peak curtailment with off-peak make-up subject to optional ramp bounds). Unless noted otherwise, we use $\rho=0.8$, on-peak $t\in\{16,17,18,19\}$, and the curtailment, make-up depths, and ramp bounds specified in Methods.

\subsection{Stage~1: Reliability pre-screening}\label{sec:stage1-results}
\begin{figure}[hbt!]
    \centering
    \includegraphics[width=0.9\textwidth]{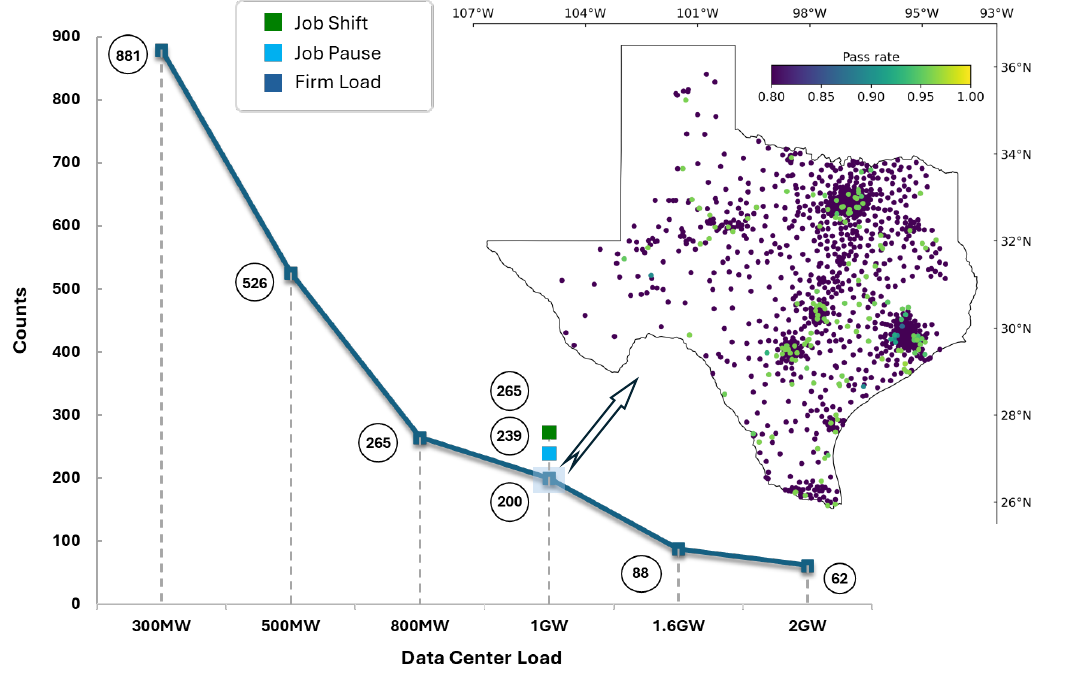}
    \caption{Stage~1: Reliability pre-screening. Qualified counts by size and envelope ($\tau_{\mathrm{PR}}=0.95$). The heat map (1~GW example) shows $\mathrm{PR}_{i,f}$ by bus; yellow indicates higher pass rates. Under the baseline envelopes, flexible operation (pause and shift) expands the feasible set relative to firm. Source data are provided as a Source Data file.}
    \label{fig:qualifiedcounts}
\end{figure}

For each candidate bus $i$ and envelope $f$, we define $\mathrm{PR}_{i,f}$ as the fraction of single-contingency DC-OPF cases that remain feasible when a data-center load trajectory $d^{(f)}_{\tau(t)}$ is added at $i$ over the study period. Buses with $\mathrm{PR}_{i,f}\ge \tau_{\mathrm{PR}}$ (here $\tau_{\mathrm{PR}}=0.95$) form the Stage~1 pre-qualified set $\mathcal{Q}^{(1)}_f$ for envelope $f$. This metric transparently summarizes location-level robustness to line or generator outages under the assumed operating envelope. We restrict to 24--500~kV interconnection points to reflect practical siting.

Importantly, Stage~1 is a reliability gate based on DC-OPF screening, while Stage~2 serves as an operational feasibility validation under SCUC/SCED. As a result, some buses in $\mathcal{Q}^{(1)}_f$ may be removed in Stage~2 if the full market simulation becomes infeasible or fails to converge.

Screening outcomes are highly uneven across sizes and envelopes, as illustrated in Figure~\ref{fig:qualifiedcounts}. At moderate sizes (300~MW and below), many locations satisfy the Stage~1 gate; at 1--2~GW, the pre-qualified set $\mathcal{Q}^{(1)}_f$ contracts sharply toward backbone corridors and generation hubs. Under the baseline envelopes used here (pause with $\alpha=0.15$ and shift with $\alpha=0.20$), flexible operation expands $\mathcal{Q}^{(1)}_f$ relative to firm, consistent with the idea that pre-agreed envelopes create headroom under contingencies. We note, however, that for some parameterizations, for example small $\alpha$ under shift, flexibility benefits can be muted or offset by make-up timing and ramp constraints, which we further assess in the sensitivity analysis.

\subsection{Stage~2: Price--congestion impact quantification}\label{sec:stage2-results}
Having secured a set of locations that satisfy physical reliability gates in Stage~1, the next critical question is whether these candidates can operate without distorting system-wide market prices or triggering operational infeasibility. To address this, for each envelope $f$ and each pre-qualified bus $i\in\mathcal{Q}^{(1)}_f$ from Stage~1, we run day-ahead security constrained unit commitment (SCUC) and intra-day security constrained economic dispatch (SCED) with the corresponding load trajectory $\ell^{(f)}_t$ placed at $i$. Stage~2 therefore acts as an operational feasibility validation: if the SCUC/SCED problem becomes infeasible or fails to converge for a given $(i,f)$, we exclude that bus from further evaluation. The remaining buses form the Stage~2 qualified set $\mathcal{Q}^{(2)}_f$, which is the final feasibility-vetted candidate pool used for reporting market impacts and for Stage~3 ranking.

We summarize system-wide absolute market outcomes using five metrics (definitions in Methods, Stage~2): (1) cross-node mean LMP, (2) cross-node LMP dispersion (P95--P5 range), (3) cross-node LMP standard deviation, (4) binding hours, and (5) congestion rent. We report medians across $i\in\mathcal{Q}^{(2)}_f$ with Interquartile Range (IQR) in brackets; the corresponding sample sizes are reported as $N_f=|\mathcal{Q}^{(2)}_f|$ in Tables~\ref{tab:flex-abs-1gw}--\ref{tab:flex-abs-2gw}.

\paragraph{Key findings}
(i) Flexibility expands the siting frontier after feasibility validation. After Stage~1 pre-screening and Stage~2 operational feasibility validation, the final qualified set sizes are as follows: at 1~GW, $N_{\mathrm{firm}}=193$, $N_{\mathrm{pause}}=226$ (+17\%), and $N_{\mathrm{shift}}=210$ (+9\%); at 2~GW, $N_{\mathrm{firm}}=57$, $N_{\mathrm{pause}}=68$ (+19\%), and $N_{\mathrm{shift}}=69$ (+21\%). The expansion is more pronounced under pause at 1~GW, while both modes show similar expansion at 2~GW.

(ii) All-hour price levels are stable, while shift lifts off-peak means slightly. Median all-hour mean LMP is essentially unchanged across envelopes (1~GW: 24.29/24.30/24.31~\$/MWh for firm/pause/shift; 2~GW: 24.32/24.32/24.32~\$/MWh for firm/pause/shift). Consistent with make-up energy, shift modestly raises off-peak means (for example, +\$0.03 at 1~GW: 23.96 $\rightarrow$ 23.99~\$/MWh; +\$0.03 at 2~GW: 23.98 $\rightarrow$ 24.01~\$/MWh), while on-peak means change little, with a slight decline under shift.

(iii) Peak-time dispersion attenuates under flexibility modes, with shift showing larger reductions. At 1~GW, on-peak P95--P5 decreases from 2.95 (firm) to 2.94 (pause; --0.3\%) and 2.90 (shift; --1.7\%)~\$/MWh; standard deviation shows similar, small declines (2.13 $\rightarrow$ 2.09/2.10). At 2~GW, on-peak P95--P5 decreases from 2.94 (firm) to 2.92 (pause; --0.7\%) and 2.84 (shift; --3.4\%)~\$/MWh; standard deviation shows similar declines (2.08 $\rightarrow$ 2.07/2.06). The reduction in peak-time dispersion is more pronounced under shift mode, particularly at higher loading. This peak-hour attenuation can come with a small increase in off-peak, and therefore all-hour, spatial dispersion under shift, consistent with make-up energy being redistributed to non-peak hours.

(iv) System-stress proxies move little. Median binding hours and congestion rents differ by at most a few hours and \$0.01--0.02~M per day across envelopes at a fixed size. At 1~GW, binding hours remain stable (112--113~h for all hours; 16--17~h on-peak; 96--97~h off-peak). All-hour congestion rent shows a slight increase under shift (0.59 $\rightarrow$ 0.61~\$M/day), while off-peak rent changes are similarly small (0.45 $\rightarrow$ 0.46~\$M/day). At 2~GW, congestion rents remain comparable across envelopes (all hours: 0.62--0.63~\$M/day; on-peak: 0.14--0.15~\$M/day; off-peak: 0.47--0.48~\$M/day).

The key takeaway from this case study is that flexibility offered by data centers enlarges the feasible siting set by \(9\text{--}17\%\) at \(1~\mathrm{GW}\) and \(19\text{--}21\%\) at \(2~\mathrm{GW}\), depending on the flexibility mode and $\alpha$, while preserving system-wide all-hour price levels. Pause mode ($\alpha=0.15$) yields the largest expansion at 1~GW (+17\%), whereas both modes expand the frontier similarly at 2~GW. Shift mode ($\alpha=0.20$) delivers the strongest reduction in peak-hour dispersion (for example, on-peak P95--P5: 2.95 $\rightarrow$ 2.90~\$/MWh at 1~GW; 2.94 $\rightarrow$ 2.84~\$/MWh at 2~GW), with only a slight off-peak mean-price lift (+\$0.03) and a small on-peak mean-price reduction. Meanwhile, congestion rents and binding hours remain broadly stable across envelopes (for example, at 1~GW, all-hour rent 0.59--0.61~\$M/day and off-peak rent 0.45--0.46~\$M/day), supporting a favorable trade-off in which flexibility unlocks additional siting options without materially worsening system-wide congestion or LMP volatility.

\subsection{Stage~3: Reliability-gated scoring yields scenario-specific shortlists}\label{sec:stage3-results}
We convert Stage~2 absolute metrics into a composite score and produce planner-ready shortlists within each envelope. Metrics are partitioned into three groups: Group~1 for peak-hour dispersion and congestion, Group~2 for off-peak side effects, and Group~3 for all-hour impacts. After direction-consistent normalization, group entropy weights capture cross-bus information content, yielding group scores $S^{\mathrm{G}g}_{i,f}$. We then apply TOPSIS to the normalized group scores to compute the closeness coefficient $\mathrm{CC}_{i,f}\in[0,1]$ (Methods, Stage~3), and use $\mathrm{CC}_{i,f}$ as the scenario-specific composite score $S_{i,f}$ to rank candidates within each envelope. We also report fixed group weights (0.70/0.20/0.10) as diagnostics, chosen a priori to prioritize peak-hour reliability and congestion in Group~1, retain penalties for off-peak side effects in Group~2, and preserve a small all-hour guardrail through Group~3, and find rankings broadly stable to these alternatives.

\paragraph{Feasible and top-$k$ fast-track locations in the Texas case study}
\begin{figure}[hbt!]
\centering
\includegraphics[width=\textwidth]{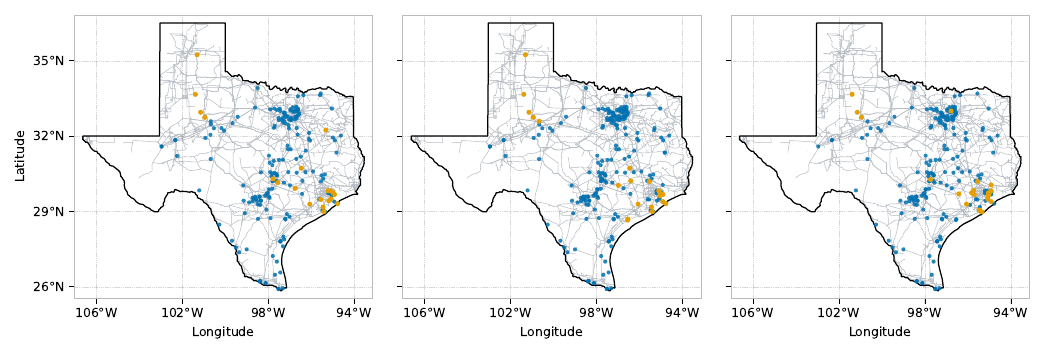}
\caption{Top-$k$ ($k=20$) locations for 1\,GW under Base, Pause, and Shift.
Amber circles denote the top-$k$ shortlist; blue circles denote other qualified locations;
gray lines show the transmission network.}
\label{fig:spatialDis-1GW}
\end{figure}

\begin{figure}[hbt!]
\centering
\includegraphics[width=\textwidth]{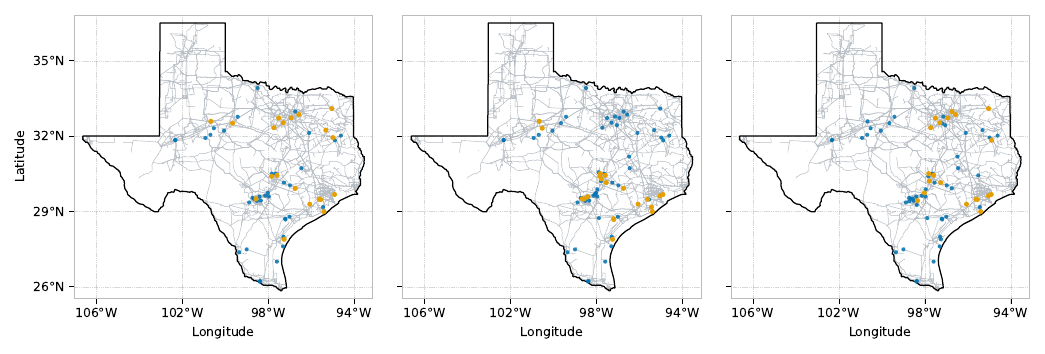}
\caption{Top-$k$ ($k=20$) locations for 2\,GW under Base, Pause, and Shift.
Amber circles denote the top-$k$ shortlist; blue circles denote other qualified locations;
gray lines show the transmission network.}
\label{fig:spatialDis-2GW}
\end{figure}

Figures~\ref{fig:spatialDis-1GW}--\ref{fig:spatialDis-2GW} illustrate the spatial distribution of Stage~2 qualified buses under each scenario, overlaid on the Texas transmission network (gray lines); amber circles denote the top-$k$ shortlist and blue circles denote the remaining qualified buses.

The maps reveal that qualified locations are not evenly spread but rather concentrate around major load hubs and backbone transmission corridors where demand is dense and the grid is most meshed. These locations, analogous to metropolitan demand centers and Gulf Coast-adjacent load zones, are precisely where new large-scale loads create both opportunities and risks due to frequent congestion. In contrast, renewable-rich zones with long radial ties contribute relatively fewer top-ranked candidates, underscoring the importance of transmission support.

Scenario sensitivity is evident at 1\,GW. The Top-20 overlap counts are Firm$\cap$Pause $=10$, Firm$\cap$Shift $=7$, and Pause$\cap$Shift $=8$. Thus, at least half of each shortlist changes with the envelope. Pause tends to surface additional Gulf Coast-adjacent and corridor nodes, whereas Shift elevates nodes in meshed northern hubs where temporal redispatch helps relieve peak congestion.

Scale dependence emerges at 2\,GW. As load grows, shortlists contract to backbone-connected hubs. Overlaps are Firm$\cap$Shift $=12$, Firm$\cap$Pause $=10$, and Pause$\cap$Shift $=7$. Firm retains more east and south-central corridor candidates, along with a few in the west; Shift tilts toward north and north-central hubs; Pause emphasizes east and south-central corridors and Gulf Coast-adjacent zones. This structure is visible in Figure~\ref{fig:spatialDis-2GW}.

\paragraph{Key findings}
(i) Flexibility not only enlarges the feasible set in Stage~1 but also reshapes the top-$k$ composition.

(ii) At 1\,GW, overlaps of only $7$--$10$ buses indicate strong envelope dependence of siting recommendations.

(iii) At 2\,GW, shortlists contract toward backbone hubs; overlaps remain non-trivial (12/10/7 across pairs), with envelope-specific tilts (Firm: east and south-central and some west; Shift: north and north-central; Pause: east and south-central and Gulf Coast-adjacent).

(iv) Overlaps are not random: common candidates concentrate in Gulf Coast-adjacent zones and northern and north-central load hubs, which appear structurally robust, while unique picks arise more often in western renewable zones or along southern and eastern corridors, where feasibility is sensitive to the assumed flexibility.

\paragraph{Interconnection time reduction via Fast Track (policy link)}
Conventional interconnection for hyperscale loads ranges from 2--6~years in favorable regions and can stretch to 5--8~years where new power supply and transmission are required, with waits of about \(\sim\!7\)~years reported in stressed hubs \cite{OnLocation2025Report,ey2025datacenter,Saul2024BloombergLawDominion,Schneider2025SiteSelection}. By contrast, planner-initiated, reliability-gated siting with pre-agreed operating envelopes has enabled first energization in 12--18~months \cite{CamusEnergy2025}. Ongoing reforms, for example SPP's proposed 90-day expedited study and ERCOT's interim large-load process, point in the same direction \cite{Plautz2025EENewsSPP,McGuireWoods2022ERCOTInterim,ERCOT2022InterimNotice,ERCOT2025LLIS}. These durations map to $T_{\mathrm{conv}}\in[5,8]$~years and $T_{\mathrm{fast}}\in[1.0,1.5]$~years in Fig.~\ref{fig:framework}, implying a conservative time-to-energization reduction of $\sim$3.5--4~years.

\paragraph{Limitations and scope}
The rankings derived in Stage~3 are based on operational metrics simulated using DC-OPF sensitivities atop SCUC/SCED on the 2000-bus synthetic Texas grid; local siting constraints such as distribution capacity, cooling and water, and parcels are exogenous. All Stage~3 quantities are reproducible from Stage~2 logs (hourly LMPs and duals), and code and inputs are portable (Methods). Importantly, while our current criteria emphasize reliability and market impacts, the workflow is designed to be extensible. As detailed in Section~2, non-electrical siting factors can be integrated as additional gates or scoring criteria in future implementations without altering the proposed workflow.

\subsection{Comparative Validation and Sensitivity Analysis}

\paragraph{Comparison against a naive historical LMP-based siting heuristic}
To benchmark the proposed framework against a common industry heuristic, we conducted a head-to-head comparison with a naive approach that selects candidate sites solely by historical average LMP, that is, last-year lowest-LMP nodes. While such price-based screening is intuitive, it does not explicitly account for $N$--1 security-constrained feasibility under large interconnection requests, nor does it quantify congestion exposure under the resulting operating conditions.

Specifically, under the base case with no added data-center load, we identified the top-20 buses with the lowest annual mean LMP as the naive market-based shortlist. We then placed a 1\,GW data center at each of these buses and evaluated them using our proposed workflow: Stage~1 (reliability gate) followed by Stage~2 (market-impact assessment) where applicable.

The results highlight a limitation of the naive approach: none of the top-20 naive candidates passed the Stage~1 reliability gate (0/20 feasible) under a 1\,GW interconnection request. Table~\ref{tab:naive-vs-proposed-top20} further shows that these historically lowest-LMP sites are geographically concentrated, primarily in the Far West, North, and South zones. This geographic concentration highlights a limitation of purely price-driven siting: low-LMP areas can coincide with locations that are sensitive to shared transmission interfaces and contingency constraints, so historical price signals alone may not translate into security-constrained interconnection headroom. By enforcing the reliability gate before ranking, the proposed framework screens out infeasible candidates and ranks sites from the qualified set, yielding shortlists that correspond to security-constrained feasibility under the study assumptions.

\paragraph{Sensitivity analysis on key parameters}
To further assess robustness beyond the baseline comparison, we conducted sensitivity analyses on key parameters, including the reliability threshold ($\tau_{\mathrm{PR}}$), flexibility depth ($\alpha$), and generation marginal costs.

\subparagraph{Impact of reliability threshold ($\tau_{\mathrm{PR}}$)}
We varied the reliability threshold to assess the stability of the reliability-gated screening. The number of qualified nodes declines monotonically as reliability requirements become more stringent, confirming that the screening mechanism behaves in a stable and physically intuitive manner. From a planning perspective, this sensitivity highlights how the threshold can function as a policy lever that trades off risk tolerance against the availability of fast-track interconnection options. Full results are provided in the Supplementary Material.

\subparagraph{Impact of flexibility depth ($\alpha$)}

We varied the curtailment and shift depth $\alpha$ to evaluate how flexibility magnitude affects system performance. Table~\ref{tab:sens-shift} summarizes results for Shift mode, where $N$ denotes the Stage~2 qualified set size among candidates that pass the Stage~1 gate. Increasing $\alpha$ expands $N$ from 188 to 253. However, the composite score ($\mathrm{CC}_{i,f}$) peaks at $\alpha=0.40$ and declines at $\alpha=0.60$, illustrating a trade-off: while deeper flexibility expands the feasible set, aggressive load shifting can increase congestion and ramp-related stress during energy make-up periods, leading to diminishing returns in the composite ranking. For small $\alpha$, for example 0.05, peak relief is often insufficient to materially improve composite-ranking outcomes.

In contrast, under Pause mode, increasing $\alpha$ slightly increases spatial price dispersion ($\mathrm{P95{-}P5}$) and does not improve the overall composite score, despite expanding the feasible set size ($N$).

\subparagraph{Robustness to other parameters}
We test robustness via uniform marginal-cost perturbation, not by altering the functional form of cost curves. Uniformly increasing generation marginal costs to simulate a 20\% fuel price shock shifts the absolute mean LMP upward but preserves the relative rank ordering of candidate sites by composite score (Spearman's rank correlation $\rho > 0.95$).

\section{Discussion}\label{sec:discussion}

Our results indicate that pre-agreed operating flexibility enlarges the feasible siting set while leaving system-wide mean price levels broadly unchanged and tempering peak-time dispersion at higher loading. This closes a gap between two mature but typically siloed practices: security-constrained feasibility studies and market-impact assessments are both well established, yet they are rarely coupled to standardized, contractable operating envelopes that can be signalled ex ante and audited ex post. By fusing a reliability gate with system-wide market metrics and a transparent multi-criteria score, the decision locus moves from bespoke, reactive studies on individual interconnection requests to a specification-driven, catalogue-style fast track.

It is important to note that the primary contribution of this work is architectural rather than exclusively so: the central novelty lies in a planner-initiated, ex ante siting workflow that integrates reliability screening, market-impact assessment, and multi-criteria ranking into a single auditable pipeline. This architectural contribution is underpinned by three methodological elements. First, the firm, pause, and shift envelopes are formally parameterized as planner-issued, contractible operating obligations rather than merely descriptive load profiles, enabling pre-certification and ex post audit. Second, the simultaneous, system-scale coupling of N--1 reliability gating with SCUC/SCED-based market simulation across multiple envelopes differs from existing hosting-capacity and interconnection studies, which are conducted on a project-by-project basis and do not jointly evaluate reliability feasibility and system-wide market impacts across a pre-defined menu of operating envelopes at the planning stage. Third, the entropy-weighted TOPSIS applied to grouped, direction-consistent absolute market metrics provides a transparent, data-driven ranking mechanism with traceable weights, a combination not previously applied in the large-load interconnection siting context. Critically, the fast-track catalogue is designed to allocate within existing grid headroom at the point of allocation, rather than clearing projects through bespoke upgrade negotiations.

As an important scope clarification, although on-site renewable generation and behind-the-meter portfolios are increasingly pursued by data-center developers, they typically do not on their own eliminate the need for grid interconnection for hyperscale, high-availability loads. Meeting firm-power requirements under low-renewable periods and contingencies would require substantial overbuild and long-duration storage, which is often less efficient in capital and reliability terms than leveraging the grid's spatial and resource diversity. Accordingly, within our workflow, on-site resources are treated as net-demand reducers, that is, as lowering the effective interconnection load net of on-site resources entering Stage~1--2, rather than as substitutes for interconnection feasibility.

Crucially, the acceleration achieved by this framework stems from procedural efficiency rather than a mere redistribution of analytical workload. Contemporary interconnection delays are dominated by wait-and-see latency: sequential queue processing, administrative idle time, and cascading re-studies triggered by speculative withdrawals, rather than pure analysis runtime. By converting this fragile, sequential dependency into a parallelized, batch-mode pre-screening process, the framework eliminates the structural dead time that constitutes the bulk of the delay. Furthermore, although this approach screens nodes that may never be utilized, the computational cost of automated batch screening in Stage~1 is parallelizable and yields durable system information products that reduce failed late-stage studies, compared to the operational and capital costs of stalled infrastructure projects under the current reactive paradigm.

Beyond the temporal dimension, the contributions of this framework carry substantive economic and societal implications that merit explicit discussion.

On the economic side, the 3.5--4 year reduction in time-to-first-energization translates directly into significant value for both load developers and the broader economy. Hyperscale data-center campuses routinely represent multi-billion-dollar capital commitments; each year of interconnection delay compounds financing costs, defers revenue, and erodes project viability. At the system level, replacing bespoke, project-by-project restudy cycles with parallelized, batch-mode pre-screening reduces the administrative overhead borne by system operators and substantially lowers the risk of cascading re-studies, a well-documented cost driver under the conventional first-come, first-served queue~\cite{ey2025datacenter,OnLocation2025Report}. More broadly, earlier energization of AI-scale compute infrastructure accelerates regional economic development and the formation of digital industry ecosystems, generating spillovers that extend well beyond the electricity sector.

On the societal side, the framework aligns with, and operationalizes, a growing policy consensus that new large loads should not impose major grid-expansion costs on existing ratepayers. The Bring Your Own New Generator (BYONG) movement, now reflected in interconnection rules across several U.S.\ jurisdictions, embodies the principle that new large loads should not impose major transmission reinforcement costs on existing ratepayers~\cite{monitoringanalytics2025,pjm2026decisional,nedc2026principles}. Analogous ratepayer-protection considerations are central to recent utility commission proceedings and legislative reviews~\cite{JLARC_2024_DataCentersVA}, and have motivated queue-reform proposals at ERCOT~\cite{ERCOT2025LLIS} and SPP~\cite{Plautz2025EENewsSPP}. While BYONG requirements represent meaningful progress, they remain fundamentally reactive and project-by-project in nature, and do not guarantee system-wide optimality.

The planner-initiated framework proposed here responds to precisely the same societal imperative, but through a proactive and system-wide lens: by pre-identifying and certifying only those locations where the existing grid, augmented by a contractible operating envelope, can accommodate new load without requiring major reinforcements, the framework structurally prevents the cost-shifting that ratepayer-protection initiatives seek to address. Fast-track eligibility is restricted exclusively to upgrade-free nodes; acceleration in interconnection therefore does not come at the expense of existing customers. In this sense, the framework operationalizes the intent behind BYONG requirements and ratepayer-protection pledges while simultaneously enabling faster, more transparent, and more equitable access to interconnection for large flexible loads. Taken together, the temporal, economic, and societal dimensions are mutually reinforcing: procedural speed is the mechanism, economic value is the proximate benefit, and ratepayer-equitable, system-operator-led planning is the broader institutional shift the framework supports.

From an energy-transition perspective, the pre-certified siting catalogue identifies locations where existing transmission headroom can accommodate large new loads without major reinforcements, the same headroom that governs the deliverability of clean power from remote renewables; co-locating data-center interconnections with renewable-accessible corridors can therefore simultaneously advance grid decarbonization and AI infrastructure deployment, with the framework's spatial outputs serving as an actionable input to joint clean-power procurement and transmission co-planning.
From a governance perspective, replacing discretionary, case-by-case interconnection decisions with a reproducible, entropy-weighted scoring mechanism reduces the risk of regulatory capture and provides a traceable basis for third-party audit and public accountability, attributes increasingly demanded by utility commissions, legislative bodies, and civil-society stakeholders as large-load interconnection volumes scale.

Building on this, the framework is readily operationalized by planners without bespoke additional tooling. Accordingly, the implementation burden shifts from bespoke, project-specific studies to standardized planner-issued specifications and a repeatable refresh process. First, planners can publish a concise menu of operating envelopes, including peak windows, curtailment and shifting depths, and optional ramp bounds, together with a public reliability gate and audit rules, so developers can design to common specifications. Next, planners can apply the Stage~1 gate at practicable interconnection levels and disclose the qualified set on a regular cadence, dynamically retiring options that become infeasible due to evolving grid topology or capacity allocation to preceding projects. This rolling update prevents stale recommendations but requires a clear final allocation rule for simultaneous claims. In practice, planners can run catalogue refreshes on a fixed cadence, for example monthly or quarterly, and additionally trigger event-driven updates after major topology changes, including new transmission entering service, major generator retirements, seasonal derates, or persistent congestion regime shifts. Each refresh replays Stage~1--3 on the latest base case, incorporates newly allocated projects as committed load additions, and republishes (i) the updated qualified set, (ii) binding corridor and substation headroom indicators, and (iii) a redacted archive of the metrics used for ranking to preserve auditability. We provide an illustrative rolling-update example in the Supplementary Material: after placing a 1~GW data center at the top-ranked node under each envelope, we update the baseline system state and re-run Stage~1--3, moving from the reliability gate to market metrics to scoring, to quantify how the qualified set, rankings, and market impacts evolve. For monitoring and refresh, rather than initial certification, periodic SCUC/SCED runs over the qualified set can quantify system-wide effects and archive logs, including LMPs and duals, for third-party review. Finally, planners can release curated shortlists defined over sites and envelopes and translate the chosen envelope into interconnection agreements with caps, telemetry, measurement and verification, and enforceable curtailment clauses so that verification replaces re-study.

To make these envelopes operationally credible, the interconnection agreement can specify: (i) a telemetry and sampling standard, with MW measured at $\Delta t$ resolution, together with a clear baseline definition; (ii) a call and response protocol, for example automated dispatch or curtailment signals with a maximum response time; (iii) measurement and verification (M\&V) rules, including allowed deviations and force-majeure clauses; and (iv) a graduated enforcement ladder, ranging from financial penalties and loss of fast-track status to physical disconnection for repeated non-compliance. These provisions convert flexibility from an aspirational attribute into an auditable operating obligation.

The workflow is portable across jurisdictions and grid contexts. The Texas synthetic grid was adopted as a fully reproducible, ISO-scale demonstration platform; it does not define the geographic scope of the framework. The same pipeline applies to any transmission system for which a network model, contingency library, and time-series demand and generation profiles are available---conditions met by most European transmission system operators, Gulf Cooperation Council system operators, and major Asian grid operators, all of which face structurally similar interconnection bottlenecks driven by the same load-growth and grid-build asymmetry documented in the Introduction. The framework relies on planning artifacts standard at most ISOs and TSOs worldwide (contingency libraries, security-constrained dispatch tools, geospatial layers), or reproducible on synthetic proxies where confidentiality precludes full model release. It is technology-agnostic across steady 24/7 compute, shiftable AI workloads, and behind-the-meter portfolios, including storage and curtailable cooling. Furthermore, the modular architecture allows planners to customize the criteria specifically to their regional priorities, for example by integrating local environmental justice screens or water constraints as discussed in the framework design, without altering the common backbone.

Methodologically, we note that the framework does not hinge on any single computational paradigm. While learning-based surrogates could accelerate repeated evaluations through screening, clustering, or warm-starting, we retain physics-based optimization, namely SCUC/SCED, as the decision core because interconnection eligibility criteria require auditability, traceable constraint satisfaction, and regulatory defensibility. We therefore position AI as a complementary acceleration layer that can be added to the workflow without changing its structure or the auditability and defensibility of the certified outcomes.

Several limitations and risks remain. The Stage~1 screen uses DC approximations and does not capture voltage or reactive-power constraints or small-signal stability; these belong to final interconnection studies. Compliance with envelopes is assumed perfect; in practice, scheduler constraints, service-level agreements, and cooling dynamics may erode effective flexibility, underscoring the need for robust M\&V design and incentives. Long-run feedbacks, including new transmission, storage, and generation induced by siting, along with distribution-interface constraints for campus-scale connections, are out of scope.

The catalogue currently evaluates sites independently and therefore does not fully model interactions among multiple simultaneous data-center sitings, including competition for shared transmission corridors, competition for substation-level headroom, and coupled congestion and ramping impacts as multiple injections reshape commitment and flows. Because multiple projects targeting the same corridor effectively compete for the same transmission headroom, deploying this framework requires a coordination mechanism for final allocation. The ranking results ($S_{i,f}$) derived in Stage~3 provide the necessary signal to implement such mechanisms. Practical pathways include: (i) score-based sequential clearing, where the planner prioritizes simultaneous requests by their composite score ($S_{i,f}$). Unlike the conventional first-come, first-served approach that processes requests blindly and often triggers extensive upgrades, this mechanism allocates only within the pre-certified, upgrade-free headroom represented by the catalogue. The highest-ranked project, that is, the one causing the least system stress, is allocated capacity first, and the system model is updated; subsequent requests are then re-evaluated against this updated baseline. To ensure transparency and mitigate strategic bunching, the planner can publish tie-break rules ex ante, for example tranche release, corridor caps, or diversification constraints. (ii) batch portfolio selection, where the planner aggregates a batch of requests and solves a joint maximization problem using the scores as welfare weights, subject to corridor-level capacity constraints and $N$--1 feasibility gates; or (iii) market-based auctions, as an optional pathway, where the planner auctions only pre-certified node--envelope capacity. The composite score can be used to qualify bidders or inform reserve prices, ensuring the allocation remains within upgrade-free headroom and bounded system impacts. Periodic simulations and data disclosures also require redaction protocols and third-party audits to balance transparency with security.

These caveats suggest a forward research agenda. Endogenous envelope design would jointly optimize envelope parameters and peak windows under welfare and verifiability constraints. Uncertainty and dynamics call for robust planning under renewable and outage uncertainty with multi-day thermal and cooling dynamics and sequential learning as fundamentals evolve. Richer physics and equity point to fast AC screens and stability checks layered onto the shortlist and to fairness weights in the multi-criteria stage to balance price-volatility relief against local impacts. Methodologically, shift-factor pre-screens and learned congestion classifiers could accelerate Stage~1 while preserving auditability.

Taken together, a planner-initiated catalogue of $N$--1 qualified locations paired with enforceable operating envelopes reframes interconnection as a structured selection among pre-vetted alternatives. In our case study, it produced transparent shortlists and traceable composite scores while moderating peak-time dispersion without materially shifting system-wide mean prices. In addition, the workflow has been exercised in a non-public setting under confidentiality constraints; however, we do not rely on those non-public results for any quantitative claims in this manuscript. Comprehensive validation on utility and ISO models and associated interconnection-queue data remains an important next step, contingent on data access and disclosure permissions. Future work will (i) validate the workflow with utility and ISO interconnection-queue data and realized timelines across multiple jurisdictions including European TSOs, Middle Eastern, and Asian grid operators, (ii) broaden the objective set beyond reliability and market metrics to include water use, acoustics, land-use and permitting, and distribution-level constraints, as well as carbon-intensity and renewable-access indicators to support energy-transition co-planning, and environmental-justice metrics alongside governance-transparency scores to enable equity-weighted, publicly auditable siting decisions at scale, and (iii) generalize envelope design under uncertainty with adaptive learning-based updates. These directions collectively support the framework's evolution toward a globally deployable planning standard for transmission-constrained systems.

\section{Methods}\label{sec:methods}

\subsection*{Data collection and processing}
To ensure the robustness of the siting framework, we constructed a dataset that captures both the spatial complexity of the transmission network and the temporal variability of operational conditions.

\subsubsection*{Data sources and topology}
The framework is demonstrated using the ACTIVSg2000 Synthetic Texas Grid, which provides a realistic topology representative of the ERCOT system \cite{Birchfield2016}. This topology includes detailed parameters for transmission lines, transformers, and thermal generators. Generator cost curves are modeled as quadratic functions with unit-specific no-load, startup, and shutdown costs. To model renewable intermittency, we integrated high-resolution 1-hour solar and wind time-series profiles sourced from the NREL Integration Dataset (Texas subset) as described in \cite{Xu2020}. These profiles are mapped to the corresponding geographic coordinates of the synthetic-grid buses.

\subsubsection*{Handling non-IID characteristics}
A critical aspect of our data processing is the preservation of temporal dependencies. Power system data, specifically load demand and renewable generation, are inherently non-independent and identically distributed because of strong autocorrelation and diurnal patterns.

Unlike static screening methods that treat operating snapshots as independent samples, our framework employs a multi-period SCUC. Consequently, we process data as a continuous chronological sequence rather than randomizing time steps. Preserving this non-IID structure is essential for two reasons:
\begin{enumerate}
    \item Generator constraints. It allows the optimization model to strictly enforce inter-temporal physical constraints, including ramping limits and minimum up and down times, which are invalidated under IID assumptions.
    \item Flexibility assessment. It enables the accurate evaluation of the Shift envelope defined in Stage~2, where a data center's ability to defer load from time $t$ to time $t+k$ depends entirely on the sequential availability of system resources.
\end{enumerate}
All load and generation profiles were synchronized to the 2020 reference year to maintain consistent correlations between weather-driven renewable output and system demand.

\subsection{System configuration and modeling assumptions}\label{sec:sys-data-compute}

\paragraph{Flexibility envelopes and profiles}\label{sec:flexibility}
We formalize three standardized envelopes,
$\mathcal{F}=\{\mathrm{firm},\mathrm{pause},\mathrm{shift}\}$,
each parameterized by utilization $\rho\in(0,1]$, a peak window
$\mathcal{T}_{\mathrm{peak}}\subset\mathcal{T}_{\mathrm{day}}$,
peak-hour curtailment $\alpha\in[0,1)$, an off-peak make-up factor
$\beta\in[0,1]$, and a within-day ramp bound $\Gamma$ in MW/h.
Let $P$ denote nameplate demand, for example $P=1$~GW or $P=2$~GW.

To avoid notational conflict with the line index $\ell\in\mathcal{L}$ used in congestion dual variables, we denote the data-center load by $d^{(f)}_\tau$ in MW, where $\tau\in\mathcal{T}_{\mathrm{day}}$ is the within-day hour. For year-hours $t\in\mathcal{T}_{\mathrm{yr}}$, we map $t$ to $\tau(t)=1+((t-1)\bmod 24)$ and use $d^{(f)}_{\tau(t)}$.

The hourly load at bus $i$ under envelope $f$ is
\begin{equation}
\label{eq:flex_load}
d^{(f)}_\tau =
\begin{cases}
\rho P, & f=\mathrm{firm},\\[4pt]
(1-\alpha_{\mathrm{pause}})\rho P, & f=\mathrm{pause},\ \tau\in\mathcal{T}_{\mathrm{peak}},\\[4pt]
\rho P, & f=\mathrm{pause},\ \tau\notin\mathcal{T}_{\mathrm{peak}},\\[4pt]
(1-\alpha_{\mathrm{shift}})\rho P, & f=\mathrm{shift},\ \tau\in\mathcal{T}_{\mathrm{peak}},\\[4pt]
\rho P + \beta_{\mathrm{shift}}
\frac{\alpha_{\mathrm{shift}}\rho P\,|\mathcal{T}_{\mathrm{peak}}|}
{24-|\mathcal{T}_{\mathrm{peak}}|},
& f=\mathrm{shift},\ \tau\notin\mathcal{T}_{\mathrm{peak}},
\end{cases}
\end{equation}
subject to the ramp condition
\begin{equation}
\label{eq:flex_ramp}
\left|d^{(f)}_{\tau}-d^{(f)}_{\tau-1}\right|\le\Gamma,
\qquad \tau\ge 2.
\end{equation}
When $\beta_{\mathrm{shift}}=1$, the daily energy of shift equals that of firm. Unless specified otherwise, the study defaults are
\begin{equation}
\label{eq:flex_defaults}
\begin{aligned}
\rho &= 0.8, \\
\mathcal{T}_{\mathrm{peak}} &= \{16,17,18,19\}, \\
\alpha_{\mathrm{pause}} &= 0.15, \\
\alpha_{\mathrm{shift}} &= 0.20, \\
\beta_{\mathrm{shift}} &= 1, \\
\Gamma &= 0.2P,
\end{aligned}
\end{equation}
where $\Gamma$ represents the data center's maximum hourly ramp capability, set to 20\% of its nameplate capacity $P$.

\paragraph{Time index sets}
We use two time index sets:
\begin{equation}
\label{eq:time_sets}
\mathcal{T}_{\mathrm{day}}=\{1,\dots,24\},
\qquad
\mathcal{T}_{\mathrm{yr}}=\{1,\dots,8760\}.
\end{equation}

\paragraph{Study system parameters}
Based on the ACTIVSg2000 model \cite{Birchfield2016}, our study system comprises
$|\mathcal{N}|=2000$ buses, $|\mathcal{G}|\approx 598$ generators, and
$|\mathcal{L}|\approx 3206$ branches. Initial interconnection candidates are defined as
\begin{equation}
\label{eq:candidate_set}
\mathcal{I}
=
\left\{
i\in\mathcal{N}:\ 24~\mathrm{kV}<V_i<500~\mathrm{kV}
\right\}.
\end{equation}
Stage~1 yields the qualified set of bus--envelope pairs
$\mathcal{Q}\subseteq\mathcal{I}\times\mathcal{F}$
via the reliability pass-rate gate; only $\mathcal{Q}$ advances to subsequent stages. Unless stated otherwise, Stage~3 results report per-envelope top-$k$ shortlists with $k=20$.

\paragraph{Modeling assumptions}
All security and market simulations utilize standard DC approximations, namely lossless power flow with fixed voltage magnitudes. Consequently, reactive power, voltage-stability limits, and small-signal stability are not explicitly modeled in this screening phase. Final interconnection studies would typically require full AC power flow and stability analyses.

\paragraph{SCUC/SCED and solvers}
Day-ahead SCUC is a MILP over $T=24$ hours with binary commitments $u_{g,t}$ and startup and shutdown variables $y_{g,t}$ and $z_{g,t}$:
\begin{equation}
\label{eq:scuc_objective}
\min
\sum_{t\in\mathcal{T}_{\mathrm{day}}}
\sum_{g\in\mathcal{G}}
\left(
C_g(p_{g,t})
+
C_g^{\mathrm{NL}}u_{g,t}
+
C_g^{\mathrm{SU}}y_{g,t}
+
C_g^{\mathrm{SD}}z_{g,t}
\right).
\end{equation}
The problem is subject to unit limits, ramping, minimum up and down times, reserves, and DC network constraints. Let $\mathbf{p}^{\mathrm{inj}}_t\in\mathbb{R}^{|\mathcal{N}|}$ denote the net bus-injection vector, which equals generation minus demand and includes the envelope load $d^{(f)}_{\tau(t)}$ at bus $i$ when present. Let $\bm{\theta}_t$ denote the bus-angle vector, $\mathbf{B}$ the bus-susceptance matrix, $\mathbf{\Phi}$ the PTDF matrix, and $\mathbf{f}_t$ the line-flow vector. We use
\begin{align}
\label{eq:dc_network}
\mathbf{B}\bm{\theta}_t &= \mathbf{p}^{\mathrm{inj}}_t, \\
\mathbf{f}_t &= \mathbf{\Phi}\mathbf{p}^{\mathrm{inj}}_t, \\
-\mathbf{F}^{\max} &\le \mathbf{f}_t \le \mathbf{F}^{\max}.
\end{align}
Intra-day SCED, formulated as an LP with piecewise-linear costs or a QP with quadratic costs, fixes $u_{g,t}$ and redispatches $p_{g,t}$ hourly. The nodal LMP vector is written as
\begin{equation}
\label{eq:lmp_decomposition}
\bm{\lambda}_t
=
\lambda^{\mathrm{sys}}_t\mathbf{1}
+
\mathbf{\Phi}^{\top}
\left(
\bm{\mu}^{+}_t-\bm{\mu}^{-}_t
\right),
\end{equation}
where the scalar components $\mu^{+}_{\ell,t}\ge 0$ and $\mu^{-}_{\ell,t}\ge 0$ represent the shadow prices on the positive and negative line limits, respectively. At optimality, at most one side binds for a given line and hour.

\paragraph{Computing environment}
All simulations were implemented in MATLAB R2023b using MATPOWER/MOST~8.0 for power-system modeling and Gurobi~9.5.1 as the optimization solver. Runs were executed on Texas A\&M University HPRC Grace nodes with 32 CPU cores and 128~GB RAM.

\subsection{Stage 1: Reliability pre-screening}\label{sec:stage1-methods}

\paragraph{Contingencies}
We consider $N-1$ outages of single branches, including lines and transformers, and individual generators; radial and island-forming elements are excluded. The contingency library $\mathcal{C}$ is fixed across sites and envelopes.

\paragraph{Pass-rate gate}
For each candidate bus $i\in\mathcal{I}$, envelope $f\in\mathcal{F}$, and year-hour $t\in\mathcal{T}_{\mathrm{yr}}$, we solve a DC-OPF under each $c\in\mathcal{C}$ with envelope-specific load $d^{(f)}_{\tau(t)}$ at bus $i$. Let $\delta_{i,f,c,t}\in\{0,1\}$ indicate feasibility, with a value of 1 when the model solves with zero penalty slack on thermal limits and 0 otherwise. The pass rate is
\begin{equation}
\label{eq:pass_rate}
\mathrm{PR}_{i,f}
=
\frac{1}{|\mathcal{T}_{\mathrm{yr}}|\,|\mathcal{C}|}
\sum_{t\in\mathcal{T}_{\mathrm{yr}}}
\sum_{c\in\mathcal{C}}
\delta_{i,f,c,t}.
\end{equation}
A site is qualified for envelope $f$ if $\mathrm{PR}_{i,f}\ge\tau_{\mathrm{PR}}$; we set $\tau_{\mathrm{PR}}=0.95$ unless stated otherwise.

\subsection{Stage 2: Market-impact simulation}\label{sec:stage2-methods}

\paragraph{Candidate load insertion}
For $(i,f)\in\mathcal{Q}$, the net demand used in SCUC and SCED is
\begin{equation}
\label{eq:load_insertion}
\widetilde{D}_{n,t}(i,f)
=
D_{n,t}
+
\mathbf{1}\{n=i\}\,d^{(f)}_{\tau(t)},
\qquad
t\in\mathcal{T}_{\mathrm{yr}}.
\end{equation}

\paragraph{Binding criterion}
Year-hour $t$ is binding for scenario $(i,f)$ if
\begin{equation}
\label{eq:binding_criterion}
\max_{\ell\in\mathcal{L}}
\left(
\mu^{(i,f)+}_{\ell,t}+\mu^{(i,f)-}_{\ell,t}
\right)
>
\varepsilon_{\mu},
\qquad
\varepsilon_{\mu}=10^{-4}\ \$/\mathrm{MWh}.
\end{equation}

\paragraph{Absolute metrics for each scenario and time window}
For a candidate bus $i\in\mathcal{Q}$ with flexibility mode $f$, let $\lambda^{(i,f)}_{n,t}$ denote the resulting LMP at bus $n\in\mathcal{N}$ and year-hour $t$ when a data center is placed at bus $i$. Define the cross-bus average LMP at hour $t$ under scenario $(i,f)$ as
\begin{equation}
\label{eq:avg_lmp_hourly}
\overline{\lambda}^{(i,f)}_{t}
=
\frac{1}{|\mathcal{N}|}
\sum_{n\in\mathcal{N}}
\lambda^{(i,f)}_{n,t}.
\end{equation}
For any time window $\mathcal{W}\subseteq\mathcal{T}_{\mathrm{yr}}$, we compute the following scenario-specific metrics:
\begin{align}
\label{eq:absolute_metrics}
\overline{\lambda}(i,f;\mathcal{W})
&=
\frac{1}{|\mathcal{W}|}
\sum_{t\in\mathcal{W}}
\overline{\lambda}^{(i,f)}_{t},
\\
\Delta_{95-5}(i,f;\mathcal{W})
&=
\frac{1}{|\mathcal{W}|}
\sum_{t\in\mathcal{W}}
\left[
\mathrm{P95}_{n}\!\left(\lambda^{(i,f)}_{n,t}\right)
-
\mathrm{P5}_{n}\!\left(\lambda^{(i,f)}_{n,t}\right)
\right],
\\
\sigma_{\lambda}(i,f;\mathcal{W})
&=
\sqrt{
\frac{1}{|\mathcal{W}|-1}
\sum_{t\in\mathcal{W}}
\left(
\overline{\lambda}^{(i,f)}_{t}
-
\overline{\lambda}(i,f;\mathcal{W})
\right)^2
},
\\
B(i,f;\mathcal{W})
&=
\sum_{t\in\mathcal{W}}
\mathbf{1}
\left\{
\max_{\ell\in\mathcal{L}}
\left(
\mu^{(i,f)+}_{\ell,t}
+
\mu^{(i,f)-}_{\ell,t}
\right)
>
\varepsilon_{\mu}
\right\},
\\
R(i,f;\mathcal{W})
&=
\frac{24}{|\mathcal{W}|}
\sum_{t\in\mathcal{W}}
\sum_{\ell\in\mathcal{L}}
\left(
\mu^{(i,f)+}_{\ell,t}
+
\mu^{(i,f)-}_{\ell,t}
\right)
F^{\max}_{\ell}.
\end{align}

We report the all-hours, on-peak, and off-peak windows. For each metric
$M\in\{\overline{\lambda},\Delta_{95-5},\sigma_{\lambda},B,R\}$,
we summarize across qualified buses as median [IQR] over
$\{M(i,f;\mathcal{W}): i\in\mathcal{Q}_{f}\}$.

\subsection{Stage 3: Scoring and shortlist construction}\label{sec:stage3-methods}

\paragraph{Metric groups and scaling}
Alternatives are indexed by $r\equiv(i,f)$ for bus--envelope pairs. We partition the Stage~2 metrics into three cost-type groups, where lower values are preferred:
\begin{align}
\label{eq:metric_groups}
\mathcal{G}_{1}
&=
\left\{
\Delta_{95-5,\mathrm{peak}},
R_{\mathrm{peak}},
B_{\mathrm{peak}},
\sigma_{\lambda,\mathrm{peak}}
\right\},
\\
\mathcal{G}_{2}
&=
\left\{
\overline{\lambda}_{\mathrm{off}},
B_{\mathrm{off}},
\sigma_{\lambda,\mathrm{off}}
\right\},
\\
\mathcal{G}_{3}
&=
\left\{
\overline{\lambda}_{\mathrm{all}},
R_{\mathrm{all}},
\sigma_{\lambda,\mathrm{all}}
\right\}.
\end{align}
Group~1 emphasizes congestion-driven price dispersion and scarcity during stress periods. Group~2 focuses on average energy cost and operational stability during off-peak hours, when load shifting occurs. Group~3 captures year-round cost efficiency and cumulative congestion impacts.

Within each group, raw metrics $X^{(g)}_{rj}$ are converted to benefit form through inverted min--max scaling:
\begin{equation}
\label{eq:minmax_invert}
B^{(g)}_{rj}
=
\frac{
\max_{r'}X^{(g)}_{r'j}-X^{(g)}_{rj}
}{
\max_{r'}X^{(g)}_{r'j}-\min_{r'}X^{(g)}_{r'j}+\epsilon
},
\qquad
\epsilon=10^{-10}.
\end{equation}

\paragraph{Entropy weights and group aggregation}
Let $R$ denote the number of alternatives. Using the same $\epsilon$ for numerical stability, we compute within-group proportions and entropies as
\begin{align}
\label{eq:entropy_props}
p^{(g)}_{rj}
&=
\frac{B^{(g)}_{rj}}{\sum_{r'}B^{(g)}_{r'j}+\epsilon},
\\
e^{(g)}_{j}
&=
-\frac{1}{\ln R}
\sum_{r}
p^{(g)}_{rj}\,
\ln\!\left(p^{(g)}_{rj}+\epsilon\right).
\end{align}
The dispersion, normalized weights, and group score are then
\begin{align}
\label{eq:entropy_weights}
d^{(g)}_{j}
&=
1-e^{(g)}_{j},
\\
w^{(g)}_{j}
&=
\frac{d^{(g)}_{j}}{\sum_{k}d^{(g)}_{k}},
\\
S^{\mathrm{G}g}_{r}
&=
\sum_{j}w^{(g)}_{j}B^{(g)}_{rj}.
\end{align}
We then stack and normalize the three group scores:
\begin{align}
\label{eq:group_vector}
\mathbf{g}_{r}
&=
\left[
S^{\mathrm{G}1}_{r},
S^{\mathrm{G}2}_{r},
S^{\mathrm{G}3}_{r}
\right],
\\
\widehat{g}_{rg}
&=
\frac{
g_{rg}
}{
\sqrt{\sum_{r'}g_{r'g}^{\,2}+\epsilon}
}.
\end{align}

\paragraph{TOPSIS on group scores}
Define the ideal best and ideal worst in the normalized group space as
\begin{align}
\label{eq:topsis_ideal}
v^{+}_{g}
&=
\max_{r}\widehat{g}_{rg},
\\
v^{-}_{g}
&=
\min_{r}\widehat{g}_{rg}.
\end{align}
The separation measures and closeness coefficient are
\begin{align}
\label{eq:topsis_cc}
S^{+}_{r}
&=
\sqrt{
\sum_{g}
\left(
\widehat{g}_{rg}-v^{+}_{g}
\right)^2
},
\\
S^{-}_{r}
&=
\sqrt{
\sum_{g}
\left(
\widehat{g}_{rg}-v^{-}_{g}
\right)^2
},
\\
\mathrm{CC}_{r}
&=
\frac{S^{-}_{r}}{S^{+}_{r}+S^{-}_{r}}
\in[0,1].
\end{align}
We use $\mathrm{CC}_{r}$ as the scenario-specific composite score $S_{i,f}$ and rank $r$ within each envelope.

\paragraph{Policy-consistent fixed group weights for diagnostics}
To complement data-driven entropy weights, we also report a policy-motivated fixed triad
\begin{equation}
\label{eq:fixed_weights}
\mathbf{w}^{\mathrm{fix}}
=
\left(
w^{\mathrm{fix}}_{1},
w^{\mathrm{fix}}_{2},
w^{\mathrm{fix}}_{3}
\right)
=
(0.70,\,0.20,\,0.10),
\end{equation}
selected a priori to mirror planning priorities: peak-hour reliability and congestion risks dominate reinforcement needs and scarcity costs; off-peak side effects warrant penalties but should not outweigh peak risk; and a modest all-hour guardrail supports year-round operational robustness. As a sensitivity diagnostic, not used for ranking, we form the weighted sum
\begin{equation}
\label{eq:fixed_score}
S^{\mathrm{fix}}_{r}
=
\sum_{g=1}^{3}
w^{\mathrm{fix}}_{g}\widehat{g}_{rg},
\end{equation}
and assess shortlist stability against the entropy-based $\mathrm{CC}_{r}$ and a uniform $(1/3,1/3,1/3)$ benchmark, for example through Spearman rank correlation and top-$k$ overlap. The primary rankings and shortlists are always based on $\mathrm{CC}_{r}$.

\paragraph{Extensibility to additional criteria}
The entropy-weighted TOPSIS framework is explicitly modular. While this study focuses on power-system metrics, the decision matrix can be extended to $M$ groups without altering the core ranking algorithm. New criteria columns can be appended to the decision matrix, and the entropy method automatically recalibrates weights based on the information content of these new factors, ensuring that the scoring remains data-driven even as the number of criteria increases.

\subsection*{Algorithmic workflow}
\begin{center}
\setlength{\fboxsep}{8pt}
\noindent\fbox{%
\begin{minipage}{0.97\linewidth}
\refstepcounter{boxcounter}
\textbf{Box \theboxcounter\;|\; Reliability-gated, flexibility-aware siting and ranking}\label{box:workflow}

\vspace{0.5em}

\noindent\textbf{Inputs:} network $(\mathcal{N},\mathcal{L},\mathcal{G})$, candidates $\mathcal{I}$, envelopes $\mathcal{F}$, contingency set $\mathcal{C}$, load and generation data, thresholds $(\tau_{\mathrm{PR}},\varepsilon_{\mu})$, time sets $(\mathcal{T}_{\mathrm{yr}},\mathcal{T}_{\mathrm{day}})$, and top-$k$.

\begin{enumerate}
    \item Define the within-day mapping
    $
    \tau(t)=1+((t-1)\bmod 24)
    $
    for $t\in\mathcal{T}_{\mathrm{yr}}$.

    \item Define
    $
    \mathcal{T}_{\mathrm{peak,yr}}
    =
    \{\,t\in\mathcal{T}_{\mathrm{yr}}:\tau(t)\in\mathcal{T}_{\mathrm{peak}}\,\}
    $
    and the windows
    $
    \mathcal{W}_{\mathrm{all}}=\mathcal{T}_{\mathrm{yr}},
    $
    $
    \mathcal{W}_{\mathrm{on}}=\mathcal{T}_{\mathrm{peak,yr}},
    $
    and
    $
    \mathcal{W}_{\mathrm{off}}
    =
    \mathcal{T}_{\mathrm{yr}}\setminus\mathcal{T}_{\mathrm{peak,yr}}.
    $

    \item For each $i \in \mathcal{I}$ and $f \in \mathcal{F}$:
    \begin{enumerate}
        \item Solve DC-OPF for all
        $
        (t,c)\in\mathcal{T}_{\mathrm{yr}}\times\mathcal{C}
        $
        with injection $d^{(f)}_{\tau(t)}$ at bus $i$.
        \item Compute $\mathrm{PR}_{i,f}$.
    \end{enumerate}

    \item Set
    $
    \mathcal{Q}
    \leftarrow
    \{
    (i,f)\in\mathcal{I}\times\mathcal{F}:
    \mathrm{PR}_{i,f}\ge\tau_{\mathrm{PR}}
    \}.
    $

    \item For each $(i,f)\in\mathcal{Q}$:
    \begin{enumerate}
        \item Build $\widetilde{D}_{n,t}(i,f)$ for $t\in\mathcal{T}_{\mathrm{yr}}$.
        \item Run SCUC and SCED to obtain
        $
        \{\lambda^{(i,f)}_{n,t},\mu^{(i,f)\pm}_{\ell,t}\}_{t\in\mathcal{T}_{\mathrm{yr}}}.
        $
        \item Compute the Stage 2 metrics over
        $
        \mathcal{W}\in
        \{
        \mathcal{W}_{\mathrm{all}},
        \mathcal{W}_{\mathrm{on}},
        \mathcal{W}_{\mathrm{off}}
        \}.
        $
    \end{enumerate}

    \item Group the metrics into $\mathcal{G}_{1}$, $\mathcal{G}_{2}$, and $\mathcal{G}_{3}$.

    \item Apply inverted min--max scaling to obtain $B^{(g)}_{rj}$.

    \item Compute entropy weights $w^{(g)}_{j}$ and normalized group scores $\widehat{g}_{rg}$.

    \item Apply TOPSIS across groups to obtain $\mathrm{CC}_{r}\in[0,1]$.

    \item Set $S_{i,f}\leftarrow\mathrm{CC}_{r}$.

    \item For each $f\in\mathcal{F}$, rank $\{(i,f)\in\mathcal{Q}\}$ by $S_{i,f}$ in descending order and release the top-$k$ shortlist.
\end{enumerate}
\end{minipage}%
}
\end{center}

\subsection{Quantifying interconnection-time savings}\label{sec:time-quant-method}

\paragraph{Definitions}
We distinguish Phase~1 time to first energization, $T_{\mathrm{fast}}$ and $T_{\mathrm{conv}}$, from total timelines that include local works, such as an onsite substation and related works, denoted by $T_{\mathrm{fast}}^{\mathrm{tot}}$ and $T_{\mathrm{conv}}^{\mathrm{tot}}$. Our primary savings metric is
\begin{equation}
\label{eq:delta_t}
\Delta T = T_{\mathrm{conv}} - T_{\mathrm{fast}}.
\end{equation}
Figure~\ref{fig:framework} also overlays conservative total timelines.

\paragraph{Evidence ranges and point estimates}
Policy and industry sources indicate
$T_{\mathrm{conv}}\in[5,8]$~years
for large loads requiring major reinforcements, whereas planner-initiated, reliability-gated paths with operating envelopes show
$T_{\mathrm{fast}}\in[1.0,1.5]$~years
to first energization \cite{ey2025datacenter,OnLocation2025Report,CamusEnergy2025,Plautz2025EENewsSPP,McGuireWoods2022ERCOTInterim,ERCOT2022InterimNotice,ERCOT2025LLIS,Saul2024BloombergLawDominion}. A conservative estimate uses the lower bound for $T_{\mathrm{conv}}$ and the upper bound for $T_{\mathrm{fast}}$:
\begin{equation}
\label{eq:delta_t_cons}
\Delta T_{\mathrm{cons}} = 5 - 1.5 = 3.5\ \text{years}.
\end{equation}
Fixing $T_{\mathrm{conv}}=5$~years and varying $T_{\mathrm{fast}}$ across its full range yields
\begin{equation}
\label{eq:delta_t_band}
\Delta T \in [3.5,\,4.0]\ \text{years},
\end{equation}
which is the range reported in the Abstract. A typical midpoint estimate, 6.5 versus 1.25~years, gives
\begin{equation}
\label{eq:delta_t_typ}
\Delta T_{\mathrm{typ}} \approx 5.25\ \text{years}.
\end{equation}
Endpoint combinations give
\begin{equation}
\label{eq:delta_t_full}
\Delta T \in [3.5,\,7.0]\ \text{years}.
\end{equation}
We interpret time to power as first meaningful energization in Phase~1, with staged caps consistent with the agreed envelope.

\paragraph{Clarification on physical versus procedural savings}
It is important to clarify that the claimed reduction in $T_{\mathrm{fast}}$ does not stem from ignoring physical construction timelines, but rather from an avoidance strategy enabled by flexibility. The Stage~1 reliability gate explicitly filters for locations where the existing grid topology, augmented by the agreed operating envelope, can accommodate the new load without violating security limits. By restricting the fast-track path specifically to these pre-validated, upgrade-free nodes, the framework allows projects to bypass the multi-year transmission-reinforcement phase that drives the 5--8 year duration of the conventional pathway. Thus, the savings reflect both procedural efficiency and the elimination of major physical works.

\paragraph{Reproducibility}
Our computational pipeline is fully reproducible end to end. Stage~1 produces the qualified set $\mathcal{Q}$; Stage~2 yields the hourly logs from which all market-impact metrics are computed; and Stage~3 applies the entropy-weighted TOPSIS procedure to obtain scenario-specific scores $S_{i,f}$.

Policy and process durations $T_{\mathrm{conv}}$ and $T_{\mathrm{fast}}$ are treated as exogenous, drawn from cited procedural sources; $\Delta T$ is then a deterministic function of these inputs. We report the same symbols in Figure~\ref{fig:framework} for notational consistency and provide document references to enable audit of the timing assumptions rather than numerical recomputation.

\backmatter

\section{Data Availability}
All data supporting the findings are provided with the paper and its Supplementary Information. The dataset used in this study is archived at Zenodo (\href{https://doi.org/10.5281/zenodo.17144790}{10.5281/zenodo.17144790}). Source data are provided with this paper.

\section{Code availability}
The code used in this study is available alongside the dataset at the same Zenodo record (\href{https://doi.org/10.5281/zenodo.17144790}{10.5281/zenodo.17144790}). This archive includes a short execution video demonstrating end-to-end runs.

\bibliography{references}

\section{Acknowledgments}
This work was supported in part by the Harvard Salata Institute and the 2026 Harvard Belfer Center–SEAS Joint Grant (L. X.). The views expressed in this paper are the opinion of the authors and do not reflect the views of PJM Interconnection, L.L.C. or its Board of Managers, of which Le Xie is a member.

\section{Author Contributions Statement}
L.X. conceived and designed the research; D.K. developed the methodology, performed the majority of simulations, and conducted the analysis and result synthesis; L.D. performed additional simulations and contributed to the analysis and synthesis; D.K. drafted the manuscript; D.K., L.D., and L.X. revised and approved the paper.

\section{Competing Interests Statement}
The authors declare no competing interests.

\clearpage
\section{Tables}

\begin{table*}[hbt!]
  \centering
  \caption{System-wide absolute metrics by flexibility mode at 1\,GW (median [IQR] across qualified buses). $N_{\mathrm{firm/pause/shift}}=193/226/210$.}
  \label{tab:flex-abs-1gw}
  \setlength{\tabcolsep}{3.5pt}
  \footnotesize
  \begin{tabular}{ccccc}
    \toprule
    Time window & Metric & Firm & Pause & Shift \\
    \midrule
    \multicolumn{5}{l}{All hours} \\
      & Mean LMP [\$/MWh]          & 24.29 [0.05] & 24.30 [0.04] & 24.31 [0.08] \\
      & LMP P95--P5 [\$/MWh]       & 2.14 [0.16] & 2.14 [0.19] & 2.19 [0.18] \\
      & LMP std [\$/MWh]           & 2.10 [0.05] & 2.10 [0.06] & 2.11 [0.07] \\
      & Binding hours [h]          & 112 [4] & 113 [5] & 113 [5] \\
      & Congestion rent [\$M/day]  & 0.59 [0.03] & 0.59 [0.04] & 0.61 [0.05] \\
      \addlinespace
    \multicolumn{5}{l}{On-peak (16--19)} \\
      & Mean LMP [\$/MWh]          & 25.96 [0.11] & 25.97 [0.10] & 25.85 [0.13] \\
      & LMP P95--P5 [\$/MWh]       & 2.95 [0.29] & 2.94 [0.35] & 2.90 [0.63] \\
      & LMP std [\$/MWh]           & 2.13 [0.09] & 2.09 [0.08] & 2.10 [0.07] \\
      & Binding hours [h]          & 16 [1] & 17 [1] & 16 [1] \\
      & Congestion rent [\$M/day]  & 0.15 [0.01] & 0.15 [0.01] & 0.15 [0.02] \\
      \addlinespace
    \multicolumn{5}{l}{Off-peak (others)} \\
      & Mean LMP [\$/MWh]          & 23.96 [0.04] & 23.97 [0.03] & 23.99 [0.07] \\
      & LMP P95--P5 [\$/MWh]       & 2.01 [0.17] & 2.01 [0.20] & 2.08 [0.23] \\
      & LMP std [\$/MWh]           & 2.10 [0.06] & 2.10 [0.06] & 2.11 [0.07] \\
      & Binding hours [h]          & 96 [3] & 96 [5] & 97 [5] \\
      & Congestion rent [\$M/day]  & 0.45 [0.02] & 0.45 [0.03] & 0.46 [0.05] \\
    \bottomrule
  \end{tabular}
\end{table*}

\begin{table*}[hbt!]
  \centering
  \caption{System-wide absolute metrics by flexibility mode at 2\,GW (median [IQR] across qualified buses). $N_{\mathrm{firm/pause/shift}}=57/68/69$.}
  \label{tab:flex-abs-2gw}
  \setlength{\tabcolsep}{3.5pt}
  \footnotesize
  \begin{tabular}{ccccc}
    \toprule
    Time window & Metric & Firm & Pause & Shift \\
    \midrule
    \multicolumn{5}{l}{All hours} \\
      & Mean LMP [\$/MWh]          & 24.32 [0.05] & 24.32 [0.08] & 24.32 [0.08] \\
      & LMP P95--P5 [\$/MWh]       & 2.22 [0.37]  & 2.25 [0.34]  & 2.22 [0.53] \\
      & LMP std [\$/MWh]           & 2.12 [0.13]  & 2.11 [0.11]  & 2.11 [0.15] \\
      & Binding hours [h]          & 113 [9]      & 115 [12]     & 115 [13] \\
      & Congestion rent [\$M/day]  & 0.62 [0.07]  & 0.62 [0.13]  & 0.63 [0.17] \\
      \addlinespace
    \multicolumn{5}{l}{On-peak (16--19)} \\
      & Mean LMP [\$/MWh]          & 25.96 [0.18] & 25.92 [0.12] & 25.88 [0.19] \\
      & LMP P95--P5 [\$/MWh]       & 2.94 [0.35]  & 2.92 [0.55]  & 2.84 [0.52] \\
      & LMP std [\$/MWh]           & 2.08 [0.13]  & 2.07 [0.10]  & 2.06 [0.12] \\
      & Binding hours [h]          & 16 [2]       & 16 [2]       & 16 [2] \\
      & Congestion rent [\$M/day]  & 0.15 [0.01]  & 0.15 [0.01]  & 0.14 [0.02] \\
      \addlinespace
    \multicolumn{5}{l}{Off-peak (others)} \\
      & Mean LMP [\$/MWh]          & 23.98 [0.08] & 24.00 [0.06] & 24.01 [0.07] \\
      & LMP P95--P5 [\$/MWh]       & 2.09 [0.46]  & 2.13 [0.53]  & 2.12 [0.67] \\
      & LMP std [\$/MWh]           & 2.11 [0.14]  & 2.13 [0.11]  & 2.12 [0.12] \\
      & Binding hours [h]          & 97 [7]       & 99 [10]      & 99 [11] \\
      & Congestion rent [\$M/day]  & 0.47 [0.07]  & 0.48 [0.11]  & 0.48 [0.15] \\
    \bottomrule
  \end{tabular}
\end{table*}

\begin{table}[hbt!]
  \centering
  \caption{Comparison between a naive historical LMP-based siting heuristic and the proposed framework (1\,GW). The naive shortlist consists of the 20 buses with the lowest annual mean LMP in the base case with no data-center load and is geographically concentrated primarily across FW, N, and S zones. Under the naive heuristic, 0/20 sites pass the Stage~1 reliability gate. The proposed shortlists report the top-20 ranked sites under each operating envelope.}
  \label{tab:naive-vs-proposed-top20}
  \setlength{\tabcolsep}{4pt}
  \footnotesize
  \begin{tabular}{cccc}
    \toprule
    Naive &
    \multicolumn{3}{c}{Proposed framework (top-20)} \\
    \cmidrule(lr){2-4}
    Bus (Zone)$^{\ast}$ &
    Firm (1\,GW) &
    Shift (1\,GW) &
    Pause (1\,GW) \\
    \midrule
    89 (FW) & 320  & 637  & 357 \\
    140 (N) & 357  & 1672 & 320 \\
    152 (N) & 188  & 320  & 188 \\
    153 (N) & 103  & 1539 & 103 \\
    55 (FW) & 1481 & 1750 & 302 \\
    101 (N) & 1831 & 1456 & 1713 \\
    3 (FW)  & 1690 & 1579 & 1815 \\
    22 (FW) & 1999 & 1446 & 1464 \\
    23 (FW) & 1837 & 1163 & 1481 \\
    182 (N) & 1126 & 1713 & 1113 \\
    170 (N) & 1809 & 1686 & 1775 \\
    28 (FW) & 1613 & 1690 & 1579 \\
    545 (S) & 1592 & 1238 & 1467 \\
    546 (S) & 1660 & 1720 & 1837 \\
    547 (S) & 1973 & 1729 & 1999 \\
    126 (N) & 1343 & 357  & 1152 \\
    499 (S) & 1089 & 1491 & 1660 \\
    139 (N) & 1491 & 188  & 1672 \\
    157 (N) & 1713 & 1448 & 1613 \\
    500 (S) & 1775 & 1464 & 1539 \\
    \bottomrule
    \multicolumn{4}{l}{\scriptsize $^{\ast}$Zone abbreviations: FW = Far West, N = North, S = South.} \\
  \end{tabular}
\end{table}

\begin{table}[hbt!]
  \centering
  \caption{Sensitivity to flexibility depth ($\alpha$) under \textbf{Shift} mode (1\,GW). $N$ denotes the Stage~2 qualified set size $|\mathcal{Q}^{(2)}_f|$. While $N$ increases monotonically, the composite score peaks at $\alpha=0.40$, illustrating diminishing returns as shifting depth increases.}
  \label{tab:sens-shift}
  \setlength{\tabcolsep}{5pt}
  \footnotesize
  \begin{tabular}{lcccc}
    \toprule
    \textbf{Metric (All-hours)} & $\boldsymbol{\alpha=0.05}$ & $\boldsymbol{\alpha=0.20}$ & $\boldsymbol{\alpha=0.40}$ & $\boldsymbol{\alpha=0.60}$ \\
    \midrule
    \textbf{Qualified Sites ($N$)} & \textbf{188} & \textbf{210} & \textbf{239} & \textbf{253} \\
    \midrule
    Mean LMP [\$/MWh]          & 24.31 [0.06] & 24.31 [0.08] & 24.33 [0.09] & 24.29 [0.06] \\
    LMP P95--P5 [\$/MWh]      & 2.96 [0.43]  & 2.90 [0.63]  & 2.87 [0.57]  & 2.79 [0.66]  \\
    Binding hours [h]          & 113 [6]      & 113 [5]      & 113 [4]      & 112 [5]      \\
    Congestion rent [\$M/day] & 0.61 [0.05]  & 0.61 [0.05]  & 0.60 [0.06]  & 0.60 [0.06]  \\
    \midrule
    \textbf{Composite Score ($\mathrm{CC}_{i,f}$)} & 0.637        & 0.669        & \textbf{0.774} & 0.726        \\
    \bottomrule
  \end{tabular}
\end{table}

\clearpage
\section*{Figure Captions}

Figure 1. Planner-initiated screening-to-certification workflow with indicative timelines.
Starting from a common grid model and data, planners apply a three-stage pipeline comprising Stage~1 reliability gating, Stage~2 price--congestion and operational-feasibility quantification, and Stage~3 scoring and ranking. This process produces (i) a pre-certified ranked catalogue of feasible locations and (ii) site-specific pre-agreed operating envelopes specifying maximum demand, ramp bounds, and optional pause or shift commitments.
A--D denote example developer applications selected from the planner-provided ranked catalogue and submitted under certified operating envelopes. Once the selected envelope is accepted and contractually signed, projects may proceed through a fast-track pathway, supporting earlier first energization, indicated by the lightning icon, on the order of 12--18~months.
For comparison, the conventional project-led pathway begins with unvetted requests under a first-come, first-served process. E--H denote example interconnection requests submitted without prior certification, in which developers independently choose locations and operating modes outside the ranked catalogue and certified envelopes. This pathway typically leads to energization on the order of 5--8~years when substantial new supply and transmission are required \cite{CamusEnergy2025,ey2025datacenter}.
Here, $PR_{i,f}$, $\tau_{PR}$ and $S_{i,f}$ denote the reliability-screening metric, screening threshold and ranking score, respectively; $N_{Shift}$, $N_{Firm}$ and $N_{Pause}$ denote candidate sets under shift, firm and pause operating envelopes; and $T_{fast}$ and $T_{conv}$ denote indicative times to energization for the planner-initiated and conventional pathways, respectively. Durations are illustrative and jurisdiction-dependent.

Figure 2. Stage~1: Reliability pre-screening. Qualified counts by size and envelope ($\tau_{\mathrm{PR}}=0.95$). The heat map (1~GW example) shows $\mathrm{PR}_{i,f}$ by bus; yellow indicates higher pass rates. Under the baseline envelopes, flexible operation (pause and shift) expands the feasible set relative to firm. Source data are provided as a Source Data file.

Figure 3. Top-$k$ ($k=20$) locations for 1\,GW under Base, Pause, and Shift.
Amber circles denote the top-$k$ shortlist; blue circles denote other qualified locations;
gray lines show the transmission network.

Figure 4. Top-$k$ ($k=20$) locations for 2\,GW under Base, Pause, and Shift.
Amber circles denote the top-$k$ shortlist; blue circles denote other qualified locations;
gray lines show the transmission network.

\end{document}